\documentclass[aps,prd,preprint,superscriptaddress,floatfix]{revtex4-1}

\usepackage{amsmath}
\usepackage{bm}
\usepackage{mathptmx}
\usepackage{color}
\usepackage{helvet}
\usepackage{courier}
\usepackage{mathrsfs}
\usepackage[dvips]{graphicx}
\usepackage{booktabs}
\usepackage{ulem}
\usepackage{color}
\def\slash#1{\not\!#1}

\DeclareMathAlphabet{\mathcal}{OMS}{cmsy}{m}{n}
\newcommand{\beq}{\begin{eqnarray}}
\newcommand{\eeq}{\end{eqnarray}}

\begin{document}


\title{New determination of $\mathcal{S} \mathcal{T} \langle N| \overline{q} D_{\mu} D_{\nu} q |N \rangle$ \\
based on recent experimental constraints}


\author{Philipp Gubler}
\email[]{pgubler@riken.jp}
\affiliation{ECT*, Villa Tambosi, 38123 Villazzano (Trento), Italy}

\author{Kie Sang Jeong}
\affiliation{Institute of Physics and Applied Physics, Yonsei University, Seoul 120-749, Korea}

\author{Su Houng Lee}
\affiliation{Institute of Physics and Applied Physics, Yonsei University, Seoul 120-749, Korea}


\date{\today}

\begin{abstract}
The symmetric and traceless part of the matrix element $\mathcal{S} \mathcal{T} \langle N| \overline{q} D_{\mu} D_{\nu} q |N \rangle$
can be determined from the second moment of the twist-3 parton distribution function $e(x)$. Recently, novel experimental data on
$e(x)$ have become available, which enables us to evaluate the magnitude of the above matrix element with considerably reduced
systematic uncertainties.
Based on the new experimental data, we show that $\mathcal{S} \mathcal{T} \langle N| \overline{q} D_{\mu} D_{\nu} q |N \rangle$
is likely to be at least an order of magnitude smaller than what previous model-based estimates have so far suggested.
We discuss the consequences of this observation for the analysis of deep inelastic scattering and QCD sum rules studies at finite density for the vector meson and the nucleon,
in which this matrix element is being used as an input parameter.
\end{abstract}


\maketitle


\section{\label{Intro} Introduction}
The traceless and symmetric component of the matrix element of the operator $\overline{q} D_{\mu} D_{\nu} q$ 
between a one-nucleon state, 
\begin{equation}
\begin{split}
\mathcal{S} \mathcal{T} \langle N| \overline{q} D_{\mu} D_{\nu} q |N \rangle 
\equiv & \frac{1}{2} \mathcal{S} \mathcal{T} \Bigl( \langle N| \overline{u} D_{\mu} D_{\nu} u |N \rangle + \langle N| \overline{d} D_{\mu} D_{\nu} d |N \rangle\Bigr), \\ 
\equiv & -e_2 \Bigl(p_{\mu} p_{\nu} - \frac{1}{4} g_{\mu \nu} M_N^2 \Bigr),
\end{split}
\label{eq:def}
\end{equation}
is known to play a role in the analysis of deep inelastic scattering (DIS) \cite{Lee} and in 
applications of the QCD sum rule method to finite density \cite{Jin,Choi}. In Eq.(\ref{eq:def}), 
$p_{\mu}$ stands for the four-momentum of the nucleon and $M_N$ is the 
nucleon mass. 
The value of $e_2$, however, has so far only been estimated via simple models or 
certain assumptions on the proportionality between different matrix elements \cite{Jin}. While such estimates 
may be fine for obtaining a first qualitative idea on the magnitude of $e_2$, it is far from clear whether 
they are quantitatively reliable. 

The situation now changed with the availability of new experimental data \cite{Courtoy}, which, as we will see, strongly 
constrain the value of $e_2$. This is possible due to the fact that $e_2$ is related to the second moment of the 
twist-3 distribution function $e(x)$ as shown below: 
\begin{equation}
\begin{split}
e_2 = & \; \int_0^1 dx x^2 e(x) \\ 
=& \; \frac{1}{2} \int_0^1 dx x^2 \big[e^u(x) + e^d(x) + e^{\overline{u}}(x) + e^{\overline{d}}(x) \big]. 
\label{eq:e2}
\end{split}
\end{equation}
Here, the various flavor components $e^q(x)$ are defined as \cite{Jaffe}
\begin{equation}
e^q(x) = \frac{1}{2M_N} \int \frac{\lambda}{2\pi} e^{i\lambda x} \langle N| \overline{q} (0) [0, \lambda n] q(\lambda n) |N \rangle, 
\end{equation}
with $[0, \lambda n]$ being the gauge link for making the above expression gauge invariant and $n$ a null vector with mass 
dimension $-1$. 
Through the analysis of experimental data on the beam-spin asymmetry of di-hadron semi-inclusive DIS obtained at the CLAS 
experiment at Jefferson Lab \cite{Courtoy}, it has become possible to extract a small number of data points for $e^{\mathrm{V}}(x)$, which 
is defined as follows: 
\begin{equation}
e^{\mathrm{V}}(x) = \frac{4}{9} \big[ e^{u}(x) - e^{\overline{u}}(x) \big] 
-\frac{1}{9} \big[ e^{d}(x) - e^{\overline{d}}(x) \big]. 
\label{eq:eV}
\end{equation}
Making use of some reasonable assumptions on the flavor structure of $e(x)$ and on its behavior in those $x$ regions, where no data points are 
available, will allow us to get an estimate of $e_2$. 

As a result, we find that even though the experimental uncertainties are still rather large, the data can constrain the magnitude of $e_2$ to be 
at least an order of magnitude smaller than values obtained from the previous simple estimates \cite{Jin}. This means that 
the matrix element $\mathcal{S} \mathcal{T} \langle N| \overline{q} D_{\mu} D_{\nu} q |N \rangle$ has 
been largely overestimated in the past DIS or QCD sum rule analyses at finite density. 

Indeed, applying the novel estimate of $e_2$ to the operator product expansion (OPE) of the electromagnetic current, 
which contains information on the 
spin-averaged structure functions $F_2$ and $F_L$, it is found that the above matrix element only gives a contribution of 3 \% or 
less compared to the experimentally extracted  values of the twist-4 effects in the second moments of these structure functions and can, therefore, 
be ignored at the presently available level of precision, which is 
in contrast to the conclusions of earlier studies. 
Furthermore, examining the OPE of the vector current correlator in nuclear matter, 
coupling to the $\rho$, $\omega$, and $\phi$ mesons, and, separately, the nuclear correlator in nuclear matter, 
we similarly find the relevant contributions to be small.  

The paper is organized as follows. 
After explaining how to extract the value of $e_2$ from the experimental data in Sec. \ref{Est}, we 
study its consequences in Sec. \ref{Disc}, which includes a discussion of both the OPE needed for 
analyzing DIS data and for the sum rule analyses of vector mesons and the nucleon at finite density. 
The paper is summarized and concluded in Sec. \ref{Summary}. 

\section{\label{Est} Estimation of $e_2$}
\subsection{Earlier simple estimates}
Before discussing the estimation of $e_2$ based on the newly available experimental data, we here for illustration and later 
comparison briefly describe two simple methods, that have so far been used to compute $e_2$. 

\subsubsection{Method 1}
Here, we will follow \cite{Lee} to estimate the needed matrix element.  
\begin{equation}
\begin{split} 
\langle N| \overline{q} D_{\mu} D_{\nu} q|N \rangle 
\simeq & - P^{\,q}_{\mu} P^{\,q}_{\nu}  \langle  N| \overline{q}  q |N \rangle  \\
=& - \frac{1}{36} p_{\mu} p_{\nu}\langle N|  \overline{q}  q |N \rangle. 
\end{split}
\label{eq:trial.eq}
\end{equation}
$P^{\,q}_{\mu}$ in the first line represents the average momentum of the quark $q$ in the nucleon, while 
the second line follows from the assumption that $P^{\,q}_{\mu}$ is about $1/6$ of the nucleon momentum $p_{\mu}$, 
as half of the nucleon momentum is carried by the gluons and the rest is divided evenly among the three valence quarks. 
Transforming this result into a traceless form, we get 
\begin{equation}
\mathcal{S} \mathcal{T} \langle N| \overline{q} D_{\mu} D_{\nu} q|N \rangle 
\simeq  - \frac{1}{36} \langle N|  \overline{q}  q |N \rangle  
 \Bigl(p_{\mu} p_{\nu} - \frac{1}{4} g_{\mu \nu} M_N^2 \Bigr), 
\end{equation}
which hence means: 
\begin{equation}
e_2 \simeq \frac{1}{36} \langle N|  \overline{q}  q |N \rangle.
\end{equation}
Expressing this through the quark mass $m_q = \frac{1}{2} (m_u + m_d)$ and 
the $\pi N$ sigma term $\sigma_{\pi N} = m_q \langle N|  \overline{q}  q |N \rangle$, 
we get 
\begin{equation} 
e_2 \simeq  \frac{1}{36} \frac{\sigma_{\pi N}}{m_q}. 
\end{equation}
Using $\sigma_{\pi N}=45\,\mathrm{MeV}$ \cite{Gasser} and $m_q = 3.5\,\mathrm{MeV}$ \cite{Olive}, 
we finally obtain 
\begin{equation}
e_2 \simeq 0.36. 
\label{eq:old.est.1}
\end{equation} 

\subsubsection{Method 2}
Here, we briefly recapitulate the discussion of \cite{Jin} to estimate the needed matrix elements. 
Combining Eqs.(4.43) and (4.49) of \cite{Jin}, we obtain  
\begin{equation}
\mathcal{S} \mathcal{T} \langle N| \overline{q} D_{\alpha} D_{\beta} q |N \rangle = 
\frac{4}{3} \frac{1}{M_N^2}
\Big(\langle N| \overline{q} D_{0} D_{0} q |N \rangle  - \frac{1}{8} \langle N| \overline{q} g \sigma \cdot G q | N \rangle \Big) 
 \Big( p_{\alpha} p_{\beta} - \frac{1}{4} M^2_N g_{\alpha \beta} \Big), 
\label{eq:DD2.100}
\end{equation}
where the second term follows from the fact that the two covariant derivatives do not commute. The above equation then 
immediately gives 
\begin{equation}
e_2 = - \frac{4}{3} \frac{1}{M_N^2}
\Big(\langle N| \overline{q} D_{0} D_{0} q |N \rangle  - \frac{1}{8} \langle N| \overline{q} g \sigma \cdot G q |N \rangle \Big).  
\end{equation}
Employing a bag model estimate, it was furthermore shown in \cite{Jin} that first term in the bracket on the right-hand side of the 
above equation is much smaller than the second one and can, therefore, be ignored. 

Next, using a parametrization proposed in \cite{Jin} 
\begin{equation}
\langle N| \overline{q} g \sigma \cdot G q |N \rangle \equiv 
m_0^2 \langle N| \overline{q} q |N \rangle, 
\label{eq:new}
\end{equation}
we are led to 
\begin{equation}
e_2 \simeq  \frac{1}{6} \frac{m_0^2}{M_N^2} \langle N| \overline{q} q |N \rangle = \frac{1}{6} \frac{m_0^2}{M_N^2} \frac{\sigma_{\pi N}}{m_q}. 
\end{equation} 
$m_0^2$ is believed to be about $m^2_0 \simeq 0.8 \pm 0.2\,\mathrm{GeV}^2$ in vacuum \cite{Belyaev}, 
which was assumed in \cite{Jin} to hold also for the one-nucleon state. This assumptions leads to 
\begin{equation}
e_2 \simeq 1.95, 
\label{eq:DD2.200}
\end{equation} 
which is about 5 times larger than Eq.(\ref{eq:old.est.1}) of the previous subsection. 

As we will see later, the estimate in Eq.(\ref{eq:DD2.200}) turns out to be 2 orders of magnitude larger than 
our updated value based on experimental constraints. This means that the ansatz of Eq.(\ref{eq:new}) with $m^2_0 \simeq 0.8 \pm 0.2\,\mathrm{GeV}^2$ is most 
likely an overestimation of the actual matrix element. We, therefore, caution the practitioners of QCD sum rules at finite 
density to be careful when making use of this parametrization. 

\subsection{\label{Expdata} Evaluation based on experimental data}
As already mentioned in the Introduction, the only information that we presently have from experiment, is the magnitude of $e^{\mathrm{V}}(x)$ 
[given in Eq.(\ref{eq:eV})] at a few values of $x$. This, obviously, does not suffice to determine $e_2$ completely and we, thus, have to introduce a number of assumptions 
on the relations of $e^{u}(x)$, $e^{d}(x)$ and their sea-quark counterparts. 
For this, we 
will have to rely partially on model calculations of $e(x)$. 
Specifically, these models are the bag model (BM) \cite{Jaffe}, the 
chiral quark soliton model ($\chi$QSM) \cite{Ohnishi}, and the 
spectator model (SM) \cite{Jakob}. 
To get an idea of the systematic uncertainties of these assumptions and models, we will test several versions of 
them and study their effects on $e_2$. 
For illustration we give the $e_2$ values and their respective flavor decompositions obtained from the BM, $\chi$QSM, and SM in 
Table \ref{tab:model.val}. 
\begin{table}
\renewcommand{\arraystretch}{1.5}
\begin{center}
\caption{Second moments of $e(x)$ for the various quark components, extracted from the bag model \cite{Jaffe}, the 
chiral quark soliton model \cite{Ohnishi} and the spectator model \cite{Jakob}.}
\vspace{0.2cm}
\label{tab:model.val}
\begin{tabular}{cccc}  
\hline 
 & BM & $\chi$QSM & SM  \\ \hline
\hspace*{0.3cm} $e^u_2 \times 10^2$ \hspace*{0.3cm}& \hspace*{0.4cm}8.1 \hspace*{0.4cm}&\hspace*{0.4cm} 5.3 \hspace*{0.4cm}& \hspace*{0.4cm}10.7 \hspace*{0.4cm} \\
$e^d_2 \times 10^2$ & 4.1 & 3.1 & \hspace*{-0.3cm}$-$2.2 \\
$e^{\overline{u}}_2 \times 10^2$ & 0.2 &  0.4 & - \\
$e^{\overline{d}}_2 \times 10^2$ & 0.3 &  0.3 & - \\
$e_2 \times 10^2$ &  6.3 & 4.6 & 4.3 \\
\hline
\end{tabular}
\end{center}
\end{table}

Throughout our whole study, we will assume the sea-quark effect on $e_2$ to be flavor symmetric. 
Namely, we will set 
\begin{equation}
\int_{0}^{1}dx x^2 e^{\overline{u}}(x) = \int_{0}^{1}dx x^2 e^{\overline{d}}(x) \equiv e^{\overline{u}}_2. 
\end{equation}
The violation of this flavor symmetry can be studied by both the BM and  $\chi$QSM, which show that 
it is only a very small effect (see Table \ref{tab:model.val}), which can be ignored here. 

Next, we will have to fix the relative strength of the $u$- and $d$-quark contributions to the second moment of $e^{V}(x)$. 
One could naturally expect that the two are proportional to the number of respective 
valence quarks; hence, 
\begin{equation}
e^{d}_2 \equiv \int_{0}^{1}dx x^2 e^{d}(x) =  \frac{1}{2} \int_{0}^{1}dx x^2 e^{u}(x) \equiv \frac{1}{2} e^{u}_2, 
\end{equation} 
which is satisfied with good accuracy by both the BM and $\chi$QSM. We will call this 
assumption ``Ansatz 1" in the following. 

The SM, however, seems to suggest a somewhat different picture, in which $e^{d}(x)$ shows an oscillating behavior and 
its second moment, therefore, even becomes a negative, but rather small number (see Fig. 9 of \cite{Cebulla} and Table \ref{tab:model.val}). 
We will, therefore, set as a second assumption 
\begin{equation}
\int_{0}^{1}dx x^2 e^{d}(x) =  0, 
\end{equation} 
which we call ``Ansatz 2". 

For Ansatz 1, we can rewrite the second moment of the experimentally measured 
function $e^{V}(x)$ as 
\begin{equation}
\int_{0}^{1}dx x^2 e^{V}(x) =  \frac{7}{18} e^{u}_2 - \frac{1}{3} e^{\overline{u}}_2, 
\end{equation} 
and for Ansatz 2, 
\begin{equation}
\int_{0}^{1}dx x^2 e^{V}(x) =  \frac{4}{9} e^{u}_2 - \frac{1}{3} e^{\overline{u}}_2.  
\end{equation}

As a last point, we need to fix the the ratio between the second moment of the $u$ and $\overline{u}$ quarks, 
for which we can obtain some guidance from the BM and the $\chi$QSM. For the BM, the $e^{\overline{u}}(x)$ is 
small and has some sizable strength only around $x=0$. Its second moment is, hence, negligible. On the other hand, 
the $e^{\overline{u}}(x)$ for the $\chi$QSM is larger, with $e^{\overline{u}}_2$ having the size of almost 10 \% of the 
valence value of $e^{u}_2$. We, thus, define 
\begin{equation}
\int_{0}^{1}dx x^2 e^{\overline{u}}(x) = \eta \int_{0}^{1}dx x^2 e^{u}(x), 
\label{eq:eta}
\end{equation}
and choose for the parameter $\eta$ a range of $0 \leq \eta \leq 0.10$. 

With this definition, we can finally relate the second moment of the experimentally measured 
function $e^{V}(x)$,  
\begin{equation}
\int_{0}^{1}dx x^2 e^{V}(x) \equiv e^{V}_2,   
\end{equation}
with $e^{u}_2$, $e^{\overline{u}}_2$, $e^{d}_2$ and $e^{\overline{d}}_2$. For Ansatz 1, we get 
\begin{equation}
\begin{split}
e^{u}_2 &= \frac{18}{7 - 6\eta} e^{V}_2, \\
e^{d}_2 &= \frac{9}{7 - 6\eta} e^{V}_2, \\
e^{\overline{u}}_2 = e^{\overline{d}}_2 &= \frac{18 \eta}{7 - 6\eta} e^{V}_2.
\end{split}
\label{eq:Ansatz1}
\end{equation}
The results for Ansatz 2 can meanwhile be given as follows: 
\begin{equation}
\begin{split}
e^{u}_2 &= \frac{9}{4 - 3\eta} e^{V}_2, \\
e^{d}_2 &= 0, \\
e^{\overline{u}}_2 = e^{\overline{d}}_2 &= \frac{9 \eta}{4 - 3\eta} e^{V}_2. 
\end{split}
\label{eq:Ansatz2}
\end{equation}

\begin{figure}
\includegraphics[width=11.0cm]{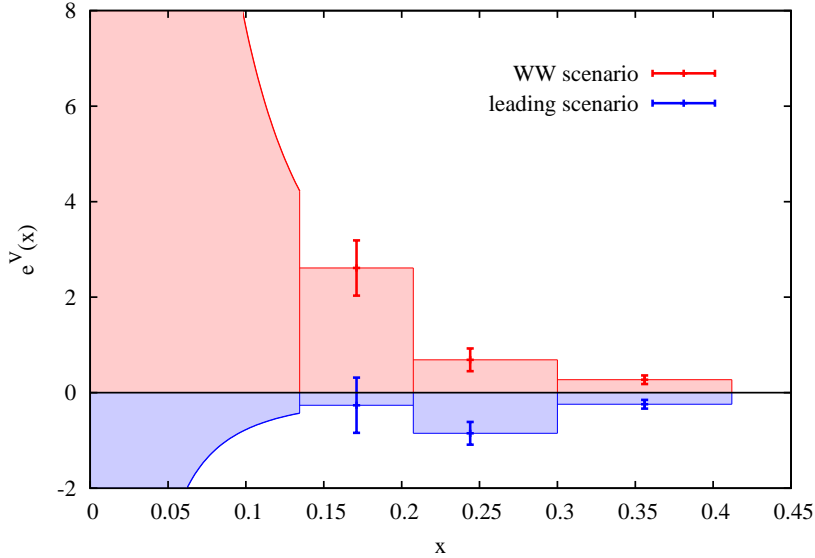}
\caption{(Color Online) The three experimentally available data points in both the WW scenario (red points) and leading scenario (blue points), 
shown together with the rectangles used for the numerical integration, as described in the text. 
In addition to the data points, the extrapolation used for $e^{V}(x)$ at small $x$, 
for which we have assumed a $1/x^2$-type form, 
 is illustrated below $x\sim0.13$ as red and blue lines, respectively.}
\label{fig:eVx}
\end{figure}

Next, we briefly discuss how to obtain $e^{V}_2$ from the experimental data of \cite{Courtoy}. 
In the region of the three data points provided by experiment, we employ the 
most simple rectangular rule to approximate the integral, as indicated in Fig. \ref{fig:eVx}. 
For applying this rule, the horizontal rectangle sizes are fixed as follows: 
1) The boundary of any two neighboring rectangles is determined to be in the center of the corresponding data points. 
2) The lowest and highest rectangles are set to be symmetric with respect to their data points. 
For the region below the lowest rectangle, we 
make use of the fact that $e^{V}(x)$ at small $x$ is known to have a 
$A/x^{1+ \alpha}$ behavior with $\alpha<1$ \cite{Jaffe}. Note that the leading 
Pomeron contribution ($\alpha=1$) vanishes here because quarks and antiquarks contribute with 
opposite signs to $e^{V}(x)$. 
However, as we 
have no additional information on the value of $\alpha$, we will 
nevertheless set $\alpha=1$ as an upper limit, fix the coefficient $A$ from the lowest data point and 
compute the respective contribution to the second moment analytically. 
Furthermore, the 
experimental points appear to quickly approach 0 above the highest data point at $x=0.356$. 
We, thus, assume it to be 0 for $x$ values above the respective highest rectangle. 
All this is pictorially illustrated in Fig.\ref{fig:eVx}, in which the shaded areas depict 
the way the numerical integration is performed. 
 
\begin{figure}
\includegraphics[width=11.0cm]{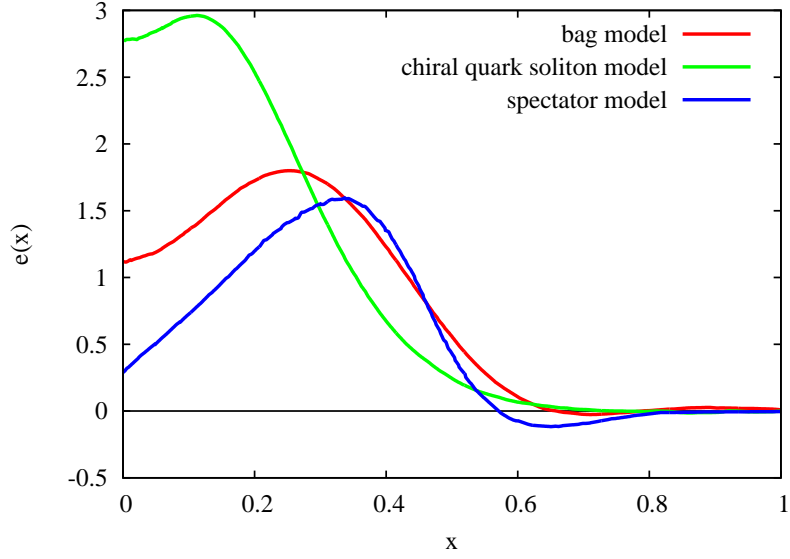}
\caption{(Color Online) The function $e(x)$ for the three models (BM, $\chi$QSM and SM) considered in this work. 
The numerical data needed for this plot have been extracted from Fig. 9 of \cite{Cebulla}.}
\label{fig:e.models}
\end{figure}

We should mention here that computing the second moment of $e(x)$ with only 
three available data points is in principle an ill-defined task. 
However, 
all of the models describing the function $e(x)$ have their 
dominant strength below 0.5, as is shown in Fig.\ref{fig:e.models}. 
Moreover, for the small-$x$ behavior of $e(x)$, we are using a conservative upper limit, which is 
based on general considerations. 
Therefore, even though we have to rely only on three data points, the values of $e(x)$ at these points can be expected to 
determine the order of magnitude of its second moment, which is all the precision needed for the present work. 

The experimental results of \cite{Courtoy}, in fact, contain the data of two different analyses. The first one is 
based on the Wandzura-Wilzcek approximation (and is, therefore, called the ``WW scenario"), while the second one includes 
terms that go beyond the Wandzura-Wilzcek approximation and is called the ``leading scenario". 
In the following, we 
will in the following consider the results of both scenarios to get a rough estimate of the systematic uncertainties 
involved. 
The numerical integration described above and depicted in Fig.\ref{fig:eVx} then gives 
\begin{align}
\int_{0}^{1}dx x^2 e^{V}(x) =
\begin{cases}
(2.35 \pm 0.32) \times 10^{-2} \,\,\,\,\,\,\,(\mathrm{WW\,\,scenario}), 
\\
(-0.97 \pm 0.32) \times 10^{-2} \,\,\,(\mathrm{leading\,\,scenario}). 
\end{cases}  
\label{eq:resultsofint}
\end{align}

All components are now in place for estimating $e_2$. First of all, using Eq.(\ref{eq:e2}) we can 
combine the various flavor contributions given in Eqs.(\ref{eq:Ansatz1}) and (\ref{eq:Ansatz2}) as 
\begin{equation}
e_2 = \frac{1}{2}(e^u_2 + e^d_2 + e^{\overline{u}}_2 + e^{\overline{d}}_2). 
\end{equation}
Applying then the results of $e^V_2$ of both the WW and leading scenario, we get altogether four values of 
$e_2$ with ranges determined by the variation of $\eta$. These ranges are shown in Table \ref{tab:exp.val} 
together with the corresponding flavor decompositions. 
\begin{table}
\renewcommand{\arraystretch}{1.5}
\begin{center}
\caption{Second moments of $e(x)$ for the various quark components, extracted from the experimental values of \cite{Courtoy}. 
The ranges in each entry of the table are obtained by employing the central values of Eq.(\ref{eq:resultsofint}) and 
varying the parameter $\eta$ of Eq.(\ref{eq:eta}) between 0 and 0.10.}
\vspace{0.2cm}
\label{tab:exp.val}
\begin{tabular}{ccccc}  
\hline 
 & \multicolumn{2}{c}{WW scenario} & \multicolumn{2}{c}{leading scenario} \\ \cmidrule(r){2-3}  \cmidrule(r){4-5} 
 & Ansatz 1 & Ansatz 2 & Ansatz 1 & Ansatz 2  \\ \hline
$e^u_2 \times 10^2$ &\hspace*{0.4cm} 6.0 $\sim$ 6.6 \hspace*{0.4cm}& \hspace*{0.4cm}5.3 $\sim$ 5.7 \hspace*{0.4cm}&
\hspace*{0.4cm} $-$2.5 $\sim$ $-$2.7 \hspace*{0.4cm}& \hspace*{0.4cm} $-$2.2 $\sim$ $-$2.4 \hspace*{0.4cm}\\
$e^d_2 \times 10^2$ & 3.0 $\sim$ 3.3 & 0.0 & $-$1.3 $\sim$ $-$1.4 & 0.0 \\
$e^{\overline{u}}_2= e^{\overline{d}}_2 \times 10^2$ \hspace{0.3cm}& 0.0 $\sim$ 0.7 & 0.0 $\sim$ 0.6 & \hspace*{0.3cm}0.0 $\sim$ $-$0.3 & \hspace*{0.3cm}0.0 $\sim$ $-$0.2  \\ 
$e_2 \times 10^2$ &  4.5 $\sim$ 5.6 & 2.7 $\sim$ 3.4 & $-$1.9 $\sim$ $-$2.3 & $-$1.1 $\sim$ $-$1.4 \\
\hline
\end{tabular}
\end{center}
\end{table}
It can be seen in this table that the largest uncertainty of $e_2$ is related to the discrepancy of the two scenarios used for analyzing 
the experimental data. 
Comparing the numbers of Table \ref{tab:exp.val} with those of Table \ref{tab:model.val}, it is observed that the model values have the same 
order of magnitude, but are generally somewhat larger than the experimentally extracted ones. 
Furthermore, it is noted that model values are mostly positive, which means that they favor the WW over the 
leading scenario (see also the recent discussion in \cite{Lorce}). 

With the four ranges obtained above, we can now give our final estimate for $e_2$. 
Taking the smallest and largest value in the bottom line of Table \ref{tab:exp.val} and, furthermore, 
allowing for the possibility that $e^V_2$ can vary in the range determined by the 
statistical errors given in Eq.(\ref{eq:resultsofint}), our result reads 
\begin{equation}
e_2 = -0.030 \sim 0.064, 
\label{eq:new.estimate.e2}
\end{equation}
which should be compared to the numbers $0.36$ of Method 1 and $1.95$ of Method 2. This comparison 
clearly shows that both Methods 1 and 2 have overestimated the relevant matrix element by at least 
an order of magnitude. Therefore, even though the experimental uncertainties of $e_2$ are 
still large, we can be quite certain that it must be much smaller than previously expected.

\section{\label{Disc} Discussion of effects on DIS and QCD sum rule analyses}

\subsection{\label{DIS} Contribution to deep inelastic electron scattering }
The relevant OPE can be obtained by using the electromagnetic current $j_{\mu}(x)=\overline{q}(x) Q \gamma_{\mu} q(x)$ in
the correlator
\begin{equation}
\Pi_{\mu\nu}(\omega,\vec{q}) = i\displaystyle \int dx^4 e^{iqx} \langle N| \mathrm{T} [j_{\mu}(x) j_{\nu}(0)] |N \rangle.
\label{eq:veccorr1}
\end{equation}
Here, $Q$ is the charge operator, and $|N \rangle$ stands for the one-nucleon state.  The twist-4 contributions, including the target mass corrections, appear as \cite{Lee}
\begin{equation}
\begin{split}
\Pi_{\mu\nu}^{\mathrm{twist}\,4}(\omega,\vec{q})=& \;
\frac{1}{x^2 Q^2} \Bigg[d_{\mu\nu}\Big( \frac{5}{8}A + \frac{1}{8}B- \frac{13}{4}C + E\Big)
+e_{\mu\nu} \Big(\frac{1}{4}A - \frac{3}{4}B - \frac{9}{2}C \Big)  \\
& \hspace{1.0cm} + e_{\mu \nu} M^2_N
\frac{q^2}{(p\cdot q)^2} \Big( \frac{1}{4}A + \frac{1}{4}B - \frac{1}{2} C +\frac{1}{2} E\Big)  \Bigg],
\end{split}
\label{eq:final.res}
\end{equation}
where we have defined
\begin{align}
e_{\mu\nu} =& \; g_{\mu\nu} - \frac{q_{\mu}q_{\nu}}{q^2}, \\
d_{\mu\nu} =& \;- \frac{p_{\mu}p_{\nu}}{(p \cdot q)^2}q^2 +(p_{\mu}q_{\nu}+p_{\nu}q_{\mu}) \frac{1}{p \cdot q} - g_{\mu\nu},
\label{eq:polarization}
\end{align}
and
\begin{align}
Q^2 =& - q^2, \\
x =& - \frac{q^2}{2(p \cdot q)}.
\label{eq:definition}
\end{align}
In the above definitions, $p^{\mu}$ stands for the four-momentum of the nucleon, with $M_N$ being the nucleon mass; hence, $p^2 = M_N^2$.
The parameters $A$-$E$ are related to the operators shown below:
\begin{equation}
\begin{split}
A_{\alpha \beta} =& \; g  \mathcal{S} \mathcal{T} \overline{q}\big[D^{\mu},G_{\beta \mu} \big]  \gamma_{\alpha} Q^2 q
= g^2 \mathcal{S} \mathcal{T} (\overline{q} \gamma_{\alpha} t^a Q^2 q) \sum_q  (\overline{q} \gamma_{\beta} t^a q) ,\\
B_{\alpha \beta} =& \; g  \mathcal{S} \mathcal{T} \overline{q}\big\{ iD_{\alpha}, \tilde{G}_{\beta \mu} \big\} \gamma^{\, \mu} \gamma_5 Q^2 q, \\
C_{\alpha \beta} =& \; m_q  \mathcal{S} \mathcal{T} \overline{q} D_{\alpha} D_{\beta} Q^2 q, \\
D_{\alpha \beta} =& \; g \mathcal{S} \mathcal{T} \overline{q}\big[D_{\alpha},G_{\mu \beta} \big]  \gamma_{\mu} Q^2 q, \\
E_{\alpha \beta} =& \; g^2 \mathcal{S} \mathcal{T} (\overline{q}t^a \gamma_5 \gamma_{\alpha} Q q ) ( \overline{q} t^a  \gamma_5 \gamma_{\beta} Q q ).
\end{split}
\label{eq:operators-dis}
\end{equation}
The expectation values of these operators are parametrized as 
\begin{equation}
\begin{split}
\langle N|A_{\alpha \beta} |N \rangle
&= \Big(p_{\alpha}p_{\beta} - \frac{1}{4} M_N^2 g_{\alpha \beta} \Big) A, \\
\langle N| B_{\alpha \beta}|N \rangle
&= \Big(p_{\alpha}p_{\beta} - \frac{1}{4} M_N^2 g_{\alpha \beta} \Big) B,\\
\langle N|C_{\alpha \beta} |N\rangle
&= \Big(p_{\alpha}p_{\beta} - \frac{1}{4} M_N^2 g_{\alpha \beta} \Big) C, \\
\langle N| D_{\alpha \beta}|N \rangle
&= \Big(p_{\alpha}p_{\beta} - \frac{1}{4} M_N^2 g_{\alpha \beta} \Big) D, \\
\langle N| E_{\alpha \beta} |N \rangle
&= \Big(p_{\alpha}p_{\beta} - \frac{1}{4} M_N^2 g_{\alpha \beta} \Big) E.
\end{split}
\label{eq:operators2}
\end{equation}
In the deep inelastic limit $Q^2 \rightarrow \infty$ and $x=$ finite, one can  neglect the target mass corrections, 
and the matrix elements contribute to the second moments of the transverse and longitudinal structure function as follows:
\begin{equation}
\begin{split}
\int_0^1 F_2^{\tau =4} dx  =& \; 0.005 \pm 0.004 ~~{\rm GeV}^2 \hfill ({\rm proton}) \\
=& \; 2M_N (\frac{1}{2} E+\frac{5}{16} A + \frac{1}{8} B- \frac{13}{8} C ),
\end{split}
\label{eq:dis1}
\end{equation}
\begin{equation}
\begin{split}
\int_0^1 F_L^{\tau =4} dx  =& \; 0.035 \pm 0.004 ~~{\rm GeV}^2 \hfill ({\rm proton}) \\
=& \; 2M_N (\frac{1}{8} A - \frac{3}{16} B- \frac{9}{4} C ).
\end{split}
\label{eq:dis2}
\end{equation}
Here, the values in the first lines are obtained from experimental constraints as extracted in \cite{Choi}.
The overall factor of $2M_N$ appearing in the second lines  comes from the covariant normalization factor of the nucleon used in extracting the above numbers.
Using the notation introduced before, the value of $C$ appearing in Eqs.(\ref{eq:dis1}) and (\ref{eq:dis2}) can be written as
\begin{equation}
\begin{split}
C =& -\frac{2}{9} m_u (e^u_2+e^{\overline{u}}_2) -\frac{1}{18} m_d (e^d_2+e^{\overline{d}}_2).
\end{split}
\label{eq:dis3}
\end{equation}
Using the values given in Table \ref{tab:exp.val} and the PDG averages for the quark masses at $1\,\mathrm{GeV}$ \cite{Olive}, 
we find that the range of  $C$ values are given as in Table \ref{tab:cval}. 
As one can see from the table, $C$ contributes less than 4\% and 1\%,  to the extracted numbers of the second moments 
for the transverse and longitudinal parts, respectively, for all cases. Hence its contribution can be safely neglected.
It should be emphasized here that
this conclusion differs from \cite{Lee}, in which an old estimate in line with Eq.(\ref{eq:old.est.1}) was used, and where the contribution of $C$ to
the second moment of the transverse structure function was found to be sizable. 
The novel findings of the present work have, therefore, somewhat changed this situation.

\begin{table}
\renewcommand{\arraystretch}{1.5}
\begin{center}
\caption{Values of $C$ that contribute to the second moments of the structure functions in units of GeV$^2$. 
The ranges in each entry of the table are obtained by employing the central values of Eq.(\ref{eq:resultsofint}) and 
varying the parameter $\eta$ of Eq.(\ref{eq:eta}) between 0 and 0.10. 
The second and third lines use factors as they appear in Eq.(\ref{eq:dis1}) and Eq.(\ref{eq:dis2}), respectively. }
\vspace{0.2cm}
\label{tab:cval}
\begin{tabular}{ccccc}
\hline
 & \multicolumn{2}{c}{WW scenario} & \multicolumn{2}{c}{leading scenario} \\ \cmidrule(r){2-3}  \cmidrule(r){4-5}
 & Ansatz 1 & Ansatz 2 & Ansatz 1 & Ansatz 2  \\ \hline
$2M_N C\times 10^5$ & \hspace*{0.4cm}$-9.9$ $  \sim$ $-12$ \hspace*{0.4cm}& \hspace*{0.4cm}$-6.8$  $\sim$ $-8.5$ \hspace*{0.4cm} &
\hspace*{0.4cm}4.1 $\sim$ 5.0 \hspace*{0.4cm}& \hspace*{0.4cm}2.8 $\sim$ 3.6 \hspace*{0.4cm}\\
$2M_N \frac{-13}{8} C\times 10^5$ & $16$ $\sim$ $20$ & \hspace*{-0.2cm}$11$ $\sim$ $14$  &  \hspace*{-0.15cm}$-$6.7$\sim$ $-8.1$&
 \hspace*{-0.1cm}$-$4.6 $\sim$ $-$5.9 \\
$2M_N \frac{-9}{4} C\times 10^5$ & $22$ $  \sim$ $27$ & \hspace*{-0.25cm} $15$  $\sim$ $19$  & \hspace*{-0.2cm} $-$9.2 $\sim$ $-$11 &
 \hspace*{-0.1cm}$-$6.3 $\sim$ $-$8.1 \\
\hline
\end{tabular}
\end{center}
\end{table}

\subsection{\label{rho} Contribution to the $\rho$ and $\omega$ meson sum rule}
The relevant OPE to the vector meson sum rule can be obtained by using the vector meson current 
$j_{\mu}(x)=\overline{q}(x) \tau \gamma_{\mu} q(x)$, where $\tau$ is the isospin operator, in
the correlator 
\begin{equation}
\Pi_{\mu\nu}(\omega,\vec{q}) = i\displaystyle \int dx^4 e^{iqx} \langle \mathrm{T} [j_{\mu}(x) j_{\nu}(0)] \rangle_{\rho}.
\label{eq:veccorr1.phi}
\end{equation}
Here, $\langle \, \rangle_{\rho}$ stands for the expectation value with respect to
the ground state of nuclear matter at $T=0$.  In the linear density approximation, the matrix element can be 
obtained as the nucleon matrix element times the density $\rho$.  Hence, the twist-4 operators contribute to the OPE
as in Eq.(\ref{eq:final.res}) multiplied by $\rho$, the only difference being that the operators in Eq.(\ref{eq:operators-dis}) 
contain the isospin matrix instead of the charge operator $Q$.  In the sum rule analysis, the vector meson is taken to be at rest $\vec{q}=0$, 
in which case there is only one invariant tensor, and the Borel transform is taken with respect to $-\omega^2  \to \infty$. 
The sum rule including the twist-4 operators, except the operator of interest here, namely $C_{\alpha \beta}$, are given 
in \cite{Hatsuda:1995dy,Leupold:1998bt}.

While the matrix elements are defined with the charge operator $Q$ replaced by the isospin operator $\tau$, the relative contribution of the operator $C_{\alpha \beta}$ relative
to the other twist-4 operators remains small as in the previous subsection, and its contribution to the vector meson sum rule can be safely neglected.  The situation could, however,
be different for the $\phi$ meson as the small $m_u,m_d$ is replaced by the larger strange quark mass $m_s$.
This case will, thus, be considered next in more detail.

\subsection{\label{Phi} The OPE for the $\phi$ meson channel in nuclear matter}
The $\phi$ meson can be described by the interpolating field $j_{\mu}(x)=\overline{s}(x) \gamma_{\mu} s(x)$,
which is substituted into the two-point function of Eq.(\ref{eq:veccorr1.phi}).
Up to dimension six and twist-2 terms, the OPE for this correlator has already been given in earlier works \cite{Hatsuda,Gubler}.
Here, we are interested in the operators of dimension six and twist-4 that contain strange quark fields.
Within the linear density approximation,
the Wilson coefficients of such operators are obtained
in analogy to the previous subsections.
The final result can then be given as Eq.(\ref{eq:final.res}) multiplied by $\rho$, the operators corresponding to
the parameters $A$-$E$ now being
\begin{equation}
\begin{split}
A_{\alpha \beta} =& \; g  \mathcal{S} \mathcal{T} \overline{s}\big[D^{\mu},G_{\beta \mu} \big]  \gamma_{\alpha} s
= g^2 \mathcal{S} \mathcal{T} (\overline{s} \gamma_{\alpha} t^a s) \sum_q  (\overline{q} \gamma_{\beta} t^a q) ,\\
B_{\alpha \beta} =& \; g  \mathcal{S} \mathcal{T} \overline{s}\big\{ iD_{\alpha}, \tilde{G}_{\beta \mu} \big\} \gamma^{\, \mu} \gamma_5 s, \\
C_{\alpha \beta} =& \; m_s  \mathcal{S} \mathcal{T} \overline{s} D_{\alpha} D_{\beta} s, \\
D_{\alpha \beta} =& \; g \mathcal{S} \mathcal{T} \overline{s}\big[D_{\alpha},G_{\mu \beta} \big]  \gamma_{\mu} s, \\
E_{\alpha \beta} =& \; g^2 \mathcal{S} \mathcal{T} (\overline{s}t^a \gamma_5 \gamma_{\alpha}s ) ( \overline{s} t^a  \gamma_5 \gamma_{\beta} s ).
\end{split}
\label{eq:operators.phi}
\end{equation}
It is noted that, as before, $D$ does not appear in the final result and we, therefore, 
do not need to be concerned with the corresponding operator any longer.
An estimate of the one-nucleon matrix elements of the other operators is given below.

\subsubsection{Numerical estimates of the twist-4 matrix elements}
We here wish to evaluate the one-nucleon matrix elements of Eq.(\ref{eq:operators.phi}).
This can, however, not be done as reliably as for the $u$- or $d$-quark case, because no experimental information from deep inelastic scattering is
available. Nevertheless, one can still try to get an estimate by making certain assumptions on the behavior of the matrix elements.
Specifically, we
will make repeated use of the ansatz 
\begin{equation}
\langle N| \overline{s} \Gamma \mathcal{O} s |N \rangle
\simeq  \langle N| \overline{u} \Gamma \mathcal{O} u |N \rangle \frac{A^s_1}{A^u_1},
\label{eq:ansatz1}
\end{equation}
for relating the $s$-quark operators with a general operator insertion $\mathcal{O}$ to their $u$-quark counterparts.
Here,  the parameters $A^{s}_1$ and $A^{u}_1$ are moments of parton distributions of the nucleon, with quark flavor $s$ and $u$. They
are defined as
\begin{align}
A^{u}_1 &= 2 \int_0^1 dx x \big[u(x) + \overline{u}(x)\big], \\
A^{s}_1 &= 2 \int_0^1 dx x \big[s(x) + \overline{s}(x)\big].
\label{eq:values2}
\end{align}
Using the recent estimation of the parton distribution functions in \cite{Martin}, the moments can be
evaluated as 
\begin{align}
A^{u}_1 &= 0.808 \pm 0.069, \\
A^{s}_1 &= 0.0443 \pm 0.0102.
\label{eq:values22}
\end{align}

\begin{itemize}
\item $\mathcal{S} \mathcal{T} \langle N|g^2 (\overline{s} \gamma_{\alpha} t^a s) \sum_q  (\overline{q} \gamma_{\beta} t^a q) |N \rangle$ \\
Using Eq.(\ref{eq:ansatz1}), we rewrite the matrix element as follows:
\begin{equation}
\mathcal{S} \mathcal{T} \langle N|g^2 (\overline{s} \gamma_{\alpha} t^a s) \sum_q  (\overline{q} \gamma_{\beta} t^a q) |N \rangle
\simeq \mathcal{S} \mathcal{T} \langle N|g^2 (\overline{u} \gamma_{\alpha} t^a u) \sum_q  (\overline{q} \gamma_{\beta} t^a q) |N \rangle \frac{A_1^s}{A_1^u}.
\label{eq:equation100}
\end{equation}
Ignoring the strange quark contribution to the sum on the right-hand side of this equation, it is seen that that it can be related to the parameter
$K^2_u$, discussed in \cite{Choi}:
\begin{equation}
\mathcal{S} \mathcal{T} \langle N|g^2 (\overline{s} \gamma_{\alpha} t^a s) \sum_q  (\overline{q} \gamma_{\beta} t^a q) |N \rangle
\simeq \frac{1}{2M_N} \big(p_{\alpha}p_{\beta} -\frac{1}{4} M_N^2 g_{\alpha \beta} \big) K^2_u \frac{A_1^s}{A_1^u}.
\label{eq:equation101}
\end{equation}
Therefore, the parameter $A$ can be given as
\begin{equation}
A = \frac{1}{2M_N} K_u^2 \frac{A_1^s}{A_1^u}.
\end{equation}

\item $\mathcal{S} \mathcal{T} \langle N|g \overline{s}\big\{ iD_{\alpha}, \tilde{G}_{\beta \mu} \big\} \gamma^{\, \mu} \gamma_5 s |N \rangle$ \\
The next matrix element can be treated in a similar way: 
\begin{equation}
\mathcal{S} \mathcal{T} \langle N| g\overline{s}\big\{ iD_{\alpha}, \tilde{G}_{\beta \mu} \big\} \gamma^{\, \mu} \gamma_5 s |N \rangle \simeq
\mathcal{S} \mathcal{T} \langle N| \overline{u}\big\{ iD_{\alpha}, \tilde{G}_{\beta \mu} \big\} \gamma^{\, \mu} \gamma_5 u |N\rangle \frac{A_1^s}{A_1^u}. 
\end{equation}
Using Eq.(19) of \cite{Choi}, which is rewritten as Eqs.(C1-3) in \cite{Kie}, 
it can be related to the parameter $K_u^g$:
\begin{equation}
\mathcal{S} \mathcal{T} \langle N| g\overline{s}\big\{ iD_{\alpha}, \tilde{G}_{\beta \mu} \big\} \gamma^{\, \mu} \gamma_5 s |N \rangle \simeq
\frac{1}{2M_N} \Big( p_{\alpha}p_{\beta} - \frac{1}{4}M^2_N g_{\alpha \beta} \Big) K_u^g \frac{A_1^s}{A_1^u}.
\end{equation}
Comparing the above result with Eq.(\ref{eq:operators2}),
we can express $B$ as 
\begin{equation}
B = \frac{1}{2M_N} K_u^g \frac{A_1^s}{A_1^u}.
\end{equation}

\item $m_s \mathcal{S} \mathcal{T} \langle N|\overline{s} D_{\alpha} D_{\beta} s |N \rangle$ \\
To study this matrix element, we will make use of the knowledge of the previous sections, which dealt with $\langle N|\overline{q} D_{\alpha} D_{\beta} q |N \rangle$.
Using again Eq.(\ref{eq:ansatz1}), we get
\begin{equation}
\begin{split}
m_s \mathcal{S} \mathcal{T} \langle N|\overline{s} D_{\alpha} D_{\beta} s |N \rangle & \simeq
m_s \mathcal{S} \mathcal{T} \langle N|\overline{u} D_{\alpha} D_{\beta} u |N \rangle \frac{A_1^s}{A_1^u} \\
&= -m_s e_2 \frac{1}{2M_N} \Big( p_{\alpha}p_{\beta} - \frac{1}{4}M^2_N g_{\alpha \beta} \Big) \frac{A_1^s}{A_1^u},
\end{split}
\end{equation}
which means that $C$ can be given as
\begin{equation}
C=-m_s e_2 \frac{A_1^s}{A_1^u}.
\label{eq:estimate.C}
\end{equation}

\item $\mathcal{S} \mathcal{T} \langle N| g^2(\overline{s}t^a \gamma_5 \gamma_{\alpha}s ) ( \overline{s} t^a  \gamma_5 \gamma_{\beta} s ) |N \rangle$ \\
The last matrix element to be determined is the one containing four strange quarks.
We use the same strategy as above and rewrite it as 
\begin{equation}
\begin{split}
\mathcal{S} \mathcal{T} \langle N| g^2(\overline{s}t^a \gamma_5 \gamma_{\alpha}s ) ( \overline{s} t^a  \gamma_5 \gamma_{\beta} s ) |N \rangle
&\simeq \mathcal{S} \mathcal{T} \langle N| g^2(\overline{s}t^a \gamma_5 \gamma_{\alpha}s ) ( \overline{u} t^a  \gamma_5 \gamma_{\beta} u) |N \rangle  \frac{A_1^s}{A_1^u} \\
&\simeq \mathcal{S} \mathcal{T} \langle N| g^2(\overline{u}t^a \gamma_5 \gamma_{\alpha}u ) ( \overline{u} t^a  \gamma_5 \gamma_{\beta} u) |N \rangle  \Big(\frac{A_1^s}{A_1^u}\Big)^2 \\
&= \frac{1}{2M_N} \Big( p_{\alpha}p_{\beta} - \frac{1}{4}M^2_N g_{\alpha \beta} \Big) \big(K_u^1 - \frac{1}{2} K_{ud}^1\big)  \Big(\frac{A_1^s}{A_1^u}\Big)^2.
\end{split}
\end{equation}
In the last line, we have used Eqs.(C1) and (C3) of \cite{Kie}.
We can, hence, express the parameter $E$ as
\begin{equation}
E= \frac{1}{2M_N} \big(K_u^1 - \frac{1}{2} K_{ud}^1\big) \Big(\frac{A_1^s}{A_1^u}\Big)^2.
\end{equation}
\end{itemize}

For extracting the values of $A$, $B$, $C$, and $E$ from the formulas of the
this subsection, we need the values of the parameters $K_u^1$, $K_u^2$,
$K_u^g$, and $K_{ud}^1$. These were estimated in \cite{Choi} from lepton-hadron deep
inelastic scattering measurement results and by assuming ratio of $K_u$ and $K_{ud}$ to
lie in a reasonable range.
We will use the findings of \cite{Choi} together with the PDG average of the strange quark mass at 1 GeV \cite{Olive} to 
give an estimate of our parameters. 
The corresponding result is given in Table \ref{tab:ABCM4}.
\begin{table}
\renewcommand{\arraystretch}{1.5}
\setlength{\tabcolsep}{10pt}
\begin{center}
\caption{Values of the parameters $A$, $B$, $C$, and $E$ in units of MeV, for six different
data sets of $K_u^1$, $K_u^2$, $K_u^g$ and $K_{ud}^1$, given in \cite{Choi}.
For $C$ and $C_{\mathrm{old}}$, we use Eq.(\ref{eq:estimate.C}) with
the central value of our new estimate of Eq.(\ref{eq:new.estimate.e2}), $e_2 \simeq 0.01$, and the old
value of Eq.(\ref{eq:DD2.200}), respectively. }
\label{tab:ABCM4}
\begin{tabular}{lccccc}
\hline
& $A$ & $B$ & $C$ & $C_{\mathrm{old}}$ & \hspace*{0.4cm}$E$ \hspace*{0.4cm} \\ \hline
$K_u^1 = K_{ud}^1/\beta$ & \hspace*{0.3cm}5.93 & $-$6.95 & $-$0.12 & $-$13.69 & \hspace*{-0.3cm}$-$1.56 \\
$K_u^1 = K_{ud}^1(\beta+1)/\beta$ & \hspace*{0.3cm}3.21 & $-$8.76 & $-$0.12 & $-$13.69 & 0.22 \\
$K_u^1 = K_{ud}^1$ & \hspace*{0.3cm}1.93 & $-$9.61 & $-$0.12 & $-$13.69 & 1.07 \\
$K_u^1 = -K_{ud}^1$ & $-$5.29 & $-$14.43 & $-$0.12 & $-$13.69 & 5.91 \\
$K_u^1 = -K_{ud}^1(\beta+1)/\beta$ & $-$6.57 & $-$15.27 & $-$0.12 & $-$13.69 & 6.76 \\
$K_u^1 = -K_{ud}^1/\beta$ & $-$9.29 & $-$17.08 & $-$0.12 &$-$13.69  & 8.54 \\
\hline
\end{tabular}
\end{center}
\end{table}

\subsubsection{Consequences for the $\phi$ meson sum rules at rest}
To get an idea of the size of the contributions of the twist-4 terms, we will here evaluate their contribution
to the OPE of the $\phi$ meson at rest. A more detailed study on the OPE for the general finite three-momentum
case will be done elsewhere.

Let us simplify Eq.(\ref{eq:final.res}) by setting $p=(M_N,0,0,0)$, which means that
we are working in the nuclear matter rest frame.
Furthermore, we set $\vec{q}=0$, take the trace over the Lorentz indices and
define
\begin{equation}
\Pi(q_0^2) = - \frac{1}{3q_0^2} \Pi^{\mu}_{\mu}(q_0,\vec{q}=0),
\end{equation}
which gives for the dimension-six, twist-4 term,
\begin{equation}
\begin{split}
\Pi^{\mathrm{dim.}\:6\mathrm{,}\,\mathrm{twist}\:4}(q_0^2) =& \; 4 M_N^2 \rho \frac{1}{q_0^6} \Big(-\frac{1}{4}A - \frac{5}{8}B - \frac{7}{4}C - \frac{1}{2} E \Big) \\
\equiv & \; 4 M_N^2 \rho \frac{1}{q_0^6} X.
\end{split}
\label{eq:phi.OPE.result}
\end{equation}
Considering the expression of $X$ together with the numbers of Table \ref{tab:ABCM4}, it is
seen that $B$ gives the dominant contribution for all data sets and that $C$ with
our new estimate is negligibly small.
It is, however, noted that with the old estimate of Eq.(\ref{eq:DD2.200}), which would mean
using $C_{\mathrm{old}}$ instead of $C$,
the $C$-term would, 
due to its large prefactor $7/4$, actually
have provided by far the largest contribution and would have increased the
overall value of $X$ at least by a factor of about 4. 
If the findings of this paper are correct, this will help to largely reduce the twist-4
contributions to this sum rule. 

Making use now of our new estimate of $C$ and of the other parameters given in Table \ref{tab:ABCM4},
the variable $X$ is found to take values that lie roughly between 4 and 9 MeV.
After Borel-transforming the result of Eq.(\ref{eq:phi.OPE.result}), we obtain
\begin{equation}
\Pi^{\mathrm{dim.}\,6\mathrm{,}\,\mathrm{twist}\,4}_{\mathrm{OPE}}(M^2,\rho) = -2 X M_N^2 \frac{\rho}{M^6}.
\label{eq:WWW}
\end{equation}
Let us compare this finding with the other terms appearing in the OPE at the same order.
These are given for instance in Eq.(12) of \cite{Gubler}, from which we list below the
linear density coefficients of the dimension-six term:
\begin{align}
&-\frac{224}{81} \pi \alpha_s  \kappa_{N} \langle \bar{s} s \rangle \langle N|\bar{s} s| N \rangle
-\frac{104}{81} m_s^3  \langle N|\bar{s} s| N \rangle
+\frac{8}{81} m_s^2 m_q \langle N|\bar{q} q| N \rangle \\
&-\frac{4}{81} m_s^2 M_N
-\frac{3}{4} m_s^2 A^s_2 M_N - \frac{5}{6} A^s_4 M_N^3.
\end{align}
Numerically, these give
\begin{align}
&1.71 \times 10^7 \,\mathrm{MeV}^3 - 5.80 \times 10^5 \,\mathrm{MeV}^3
+2.01 \times 10^4 \,\mathrm{MeV}^3 \\
&- 4.19 \times 10^5 \,\mathrm{MeV}^3
- 2.80 \times 10^5 \,\mathrm{MeV}^3 - 7.61 \times 10^5 \,\mathrm{MeV}^3,
\end{align}
for which we have used $\kappa_N=1$ and $\sigma_{sN} = m_s \langle N|\bar{s} s| N \rangle = 50\,\mathrm{MeV}$, which is rather on
the larger side considering the recent lattice QCD calculations of this quantity.
For the other parameters, we used the values given in \cite{Gubler}.
It is clear from the above numbers that the four-quark condensate term is dominant.
All this now has to be compared to Eq.(\ref{eq:WWW}), without the density and Borel mass factors. This leads
to
\begin{equation}
 -2 X M_N^2  \simeq - 1.13 \times 10^7 \,\mathrm{MeV}^3,
\label{eq:twist4}
\end{equation}
where we have set $X=6.40\,\mathrm{MeV}$, which lies at the center of the expected value range of $X$.
Even though our new estimate of the $C$ operator has helped to reduce the magnitude of the twist-4
contribution, the result of Eq.(\ref{eq:twist4}) is still rather large, namely,
of the same order as four-quark condensate term and more than 16 times larger than the twist-2 term.

\subsection{Consequence on the nucleon sum rule and nuclear symmetry energy}

The sum rule for the proton can be obtained by looking at the  correlation function of the  proton interpolating fields. 
\begin{align}
\Pi(q) &\equiv i \int d^4 x e^{iqx} \langle
\textrm{T}[\eta(x)\bar{\eta}(0)]\rangle_\rho  \nonumber\\
&=  \Pi_s(q^2,qu)+\Pi_q(q^2,qu) \slash 
q+\Pi_u(q^2,qu) \slash u, 
\label{eq:corr}
\end{align}
where $q$ is the external momentum, $u$ is the medium four-velocity
and $\Pi_s(q^2,qu)$, $\Pi_q(q^2,qu)$, and $\Pi_u(q^2,qu)$ are the three
invariants according to Lorentz, parity and time reversal symmetries
\cite{Furnstahl:1992pi}. The Ioffe current \cite{Ioffe:1983ju} will
be used as a proton interpolating field:
\begin{align}
\eta(x)=\epsilon_{abc}[u^T_a(x)C\gamma_\mu u_b(x)]\gamma_5\gamma^{\, \mu}
d_c(x).\label{crnt}
\end{align}

\subsubsection{OPE for the proton}

The OPE for proton correlator of Eq.(\ref{eq:corr}) is 
given, for instance, in \cite{Kie}. Including, furthermore, 
the $\alpha_s$
corrections \cite{Ovchinnikov:1991mu,Shiomi:1995nf,Groote:2008hz}
and the dimension-eight contributions \cite{Yang:1993bp}, it can be
summarized as follows:
\begin{align}
\Pi^{\textrm{OPE}}_s(q^2,qu)  =&
-\frac{1}{4\pi^2}(-q^2)\ln(-q^2/\mu^2)\left(
1+\frac{3}{2}\frac{\alpha_s}{\pi}\right)\langle \bar{d}d
\rangle_{\rho,I}+\frac{4}{3\pi^2}\frac{q_0^2}{q^2}\langle \bar{d}
\{iD_{0} iD_{0}\}
d \rangle_{\rho,I},\\
\Pi^{\textrm{OPE}} _q(q^2,qu) =& -\frac{1}{64
\pi^4}(-q^2)^2\ln(-q^2/\mu^2)\left(
1+\frac{71}{12}\frac{\alpha_s}{\pi}-\frac{1}{2}\frac{\alpha_s}{\pi}\ln(-q^2/\mu^2)\right)\nonumber\\
& - \frac{2}{3}
\frac{1}{q^2}\left(1-\frac{5}{6}\frac{\alpha_s}{\pi}-\frac{1}{3}\frac{\alpha_s}{\pi}\ln(-q^2/\mu^2)\right)\langle\bar{u}u
\rangle_{\rho,I}^2 \nonumber\\
&+ \frac{1}{6}\left(-\frac{1}{q^2}\right)^2\langle \bar{u} u \rangle
\langle g_s \bar{u} \sigma  \mathcal{G} u
\rangle_{\rho,I}\nonumber\\
 & + (qu) \left[ \frac{1}{6\pi^2}\ln(-q^2/\mu^2)\left(
1+\frac{7}{2}\frac{\alpha_s}{\pi}-\frac{1}{2}\frac{\alpha_s}{\pi}\ln(-q^2/\mu^2)\right)
\right](\langle u^\dagger u
\rangle_{\rho,I} + \langle d^\dagger d \rangle_{\rho,I}),\\
\Pi^{\textrm{OPE}}_u(q^2,qu)  =&
-\frac{1}{12\pi^2}(-q^2)\ln(-q^2/\mu^2)\left(
1+\frac{15}{4}\frac{\alpha_s}{\pi}-\frac{1}{2}\frac{\alpha_s}{\pi}\ln(-q^2/\mu^2)\right)(7\langle
u^\dagger u \rangle_{\rho,I} + \langle d^\dagger d
\rangle_{\rho,I}).
\end{align}
Here, $\mu$ is the 
renormalization scale which will be matched with the Borel mass $M$
after the Borel transformation. Note that the operator of interest appears as the last term in the scalar self-energy.

The density dependencies of all the operators are given in \cite{Kie} except for that of dimension eight, 
for which we use the approximation 
\begin{align}
\langle [\bar{q} q]\rangle\langle[g_s\bar{q}\sigma
\cdot\mathcal{G}q] \rangle_{\rho,I} =&~\langle [\bar{q}
q]\rangle_{\textrm{vac}} \langle[g_s\bar{q}\sigma \cdot\mathcal{G}q]
\rangle_{\textrm{vac}}\nonumber\\
&~+\big( \langle [\bar{q} q]\rangle_{\textrm{vac}}
\langle[g_s\bar{q}\sigma \cdot\mathcal{G}q]_0 \rangle_{p}+\langle
[\bar{q} q]_0\rangle_{p} \langle[g_s\bar{q}\sigma \cdot\mathcal{G}q]
\rangle_{\textrm{vac}}\big)
\left[1\mp\frac{\mathcal{R}_{-}(m_q)}{\mathcal{R}_{+}(m_q)}
I\right]\rho,
\end{align}
where $\langle[g_s\bar{q}\sigma \cdot\mathcal{G}q]
\rangle_{\textrm{vac}} = (0.8)\langle [\bar{q}
q]\rangle_{\textrm{vac}}$ and ``$+$''
 and ``$-$'' stand for $u$ and $d$ quark
flavors, respectively.
$\langle[g_s\bar{q}\sigma\cdot\mathcal{G}q]_0\rangle_{p}$ is chosen
to be $3~\textrm{GeV}^2$ as in \cite{Jin:1993up,Belyaev}. The detailed definition and
description of the ratio $\mathcal{R}_{\pm}(m_q)$ can be
found in Sec. III B of \cite{Kie}.

\subsubsection{Result for the nucleon sum rule}

The analysis for the nucleon and the symmetry energy follows that
given in \cite{Kie}, with $f=-0.3$.  Here, $f$ parametrizes the degree of factorization of four quark operator in medium $\langle \bar{q}q \rangle^2_\rho \rightarrow (1-f) \langle \bar{q}q \rangle^2_{vac}+f\langle \bar{q}q \rangle^2_\rho$.  The formalism and
other parameters are based on the nucleon sum rule at
$-q^2\rightarrow\infty$ but fixed $\vert \vec{q} \vert$
\cite{Furnstahl:1992pi}.

\begin{figure}
\includegraphics[width=5cm]{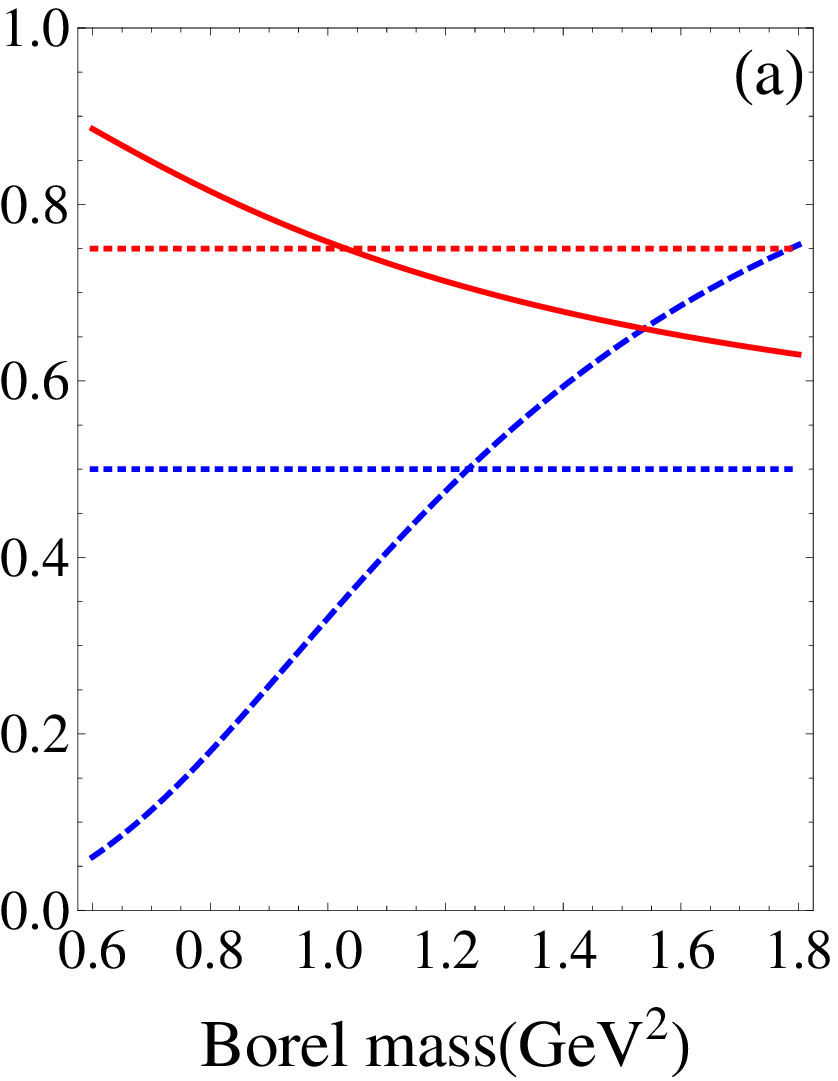}
\includegraphics[width=5cm]{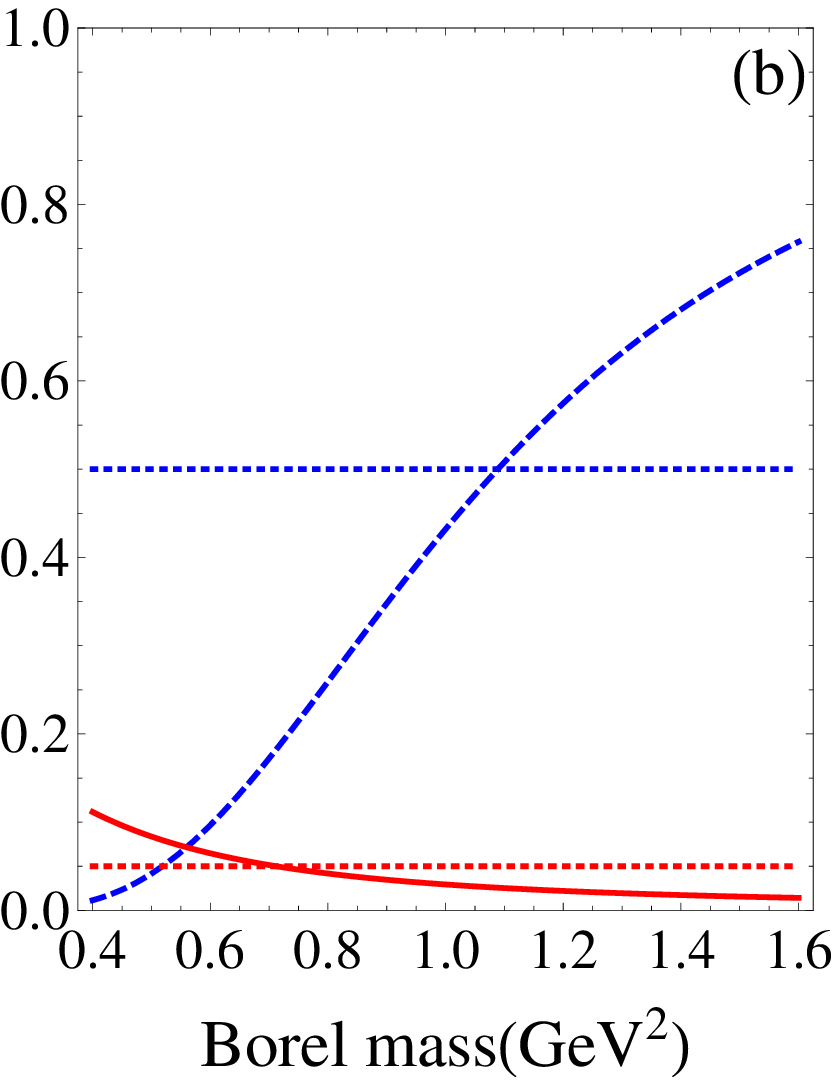}
\caption{(Color Online) Borel window for
$\bar{\mathcal{B}}[\Pi_q(q_0^2,\vert\vec{q}\vert)]$ (a) with
OPE up to dimension-six condensates and (b) dimension-eight condensates together with $\alpha_s$ corrections. 
The solid (dashed) lines show the ratios between the contribution of the highest dimensional operators (the continuum) 
and that of the total OPE. 
The horizontal dotted lines correspond to the ratios being 0.75 (red) and 0.5 (blue) in (a) and  0.05 (red) and 0.5 (blue) in (b), respectively.} 
\label{borelw}
\end{figure}

First, we show the Borel window before (a) and after (b) including the $\alpha_s$
corrections and dimension-eight condensates in Fig.~\ref{borelw}. 
The left panel of Fig (3) is adapted from \cite{Kie}. There, the upper bound for the Borel mass was obtained by requiring the ratio between the 
contribution of the highest dimensional operators and that of the total OPE to be less than 50\%, and the lower 
bound from the condition that the ratio between 
the continuum contribution and the total OPE should be less than 75\%, 
which confines the Borel window 
to 1.0 GeV$^2  <  M^2 < 1.2$ GeV$^2$.  
The added 
corrections lead to a considerable improvement, as can be inferred by 
comparing plots (a) and (b). 
We can, therefore, now require above the ratios to be less than 50\,\% and 
5\,\%, respectively, which confines the Borel window to 
$0.8~\textrm{GeV}^2\leq M^2 \leq 1.0  ~\textrm{GeV}^2$. 

\begin{figure}
\includegraphics[width=5cm]{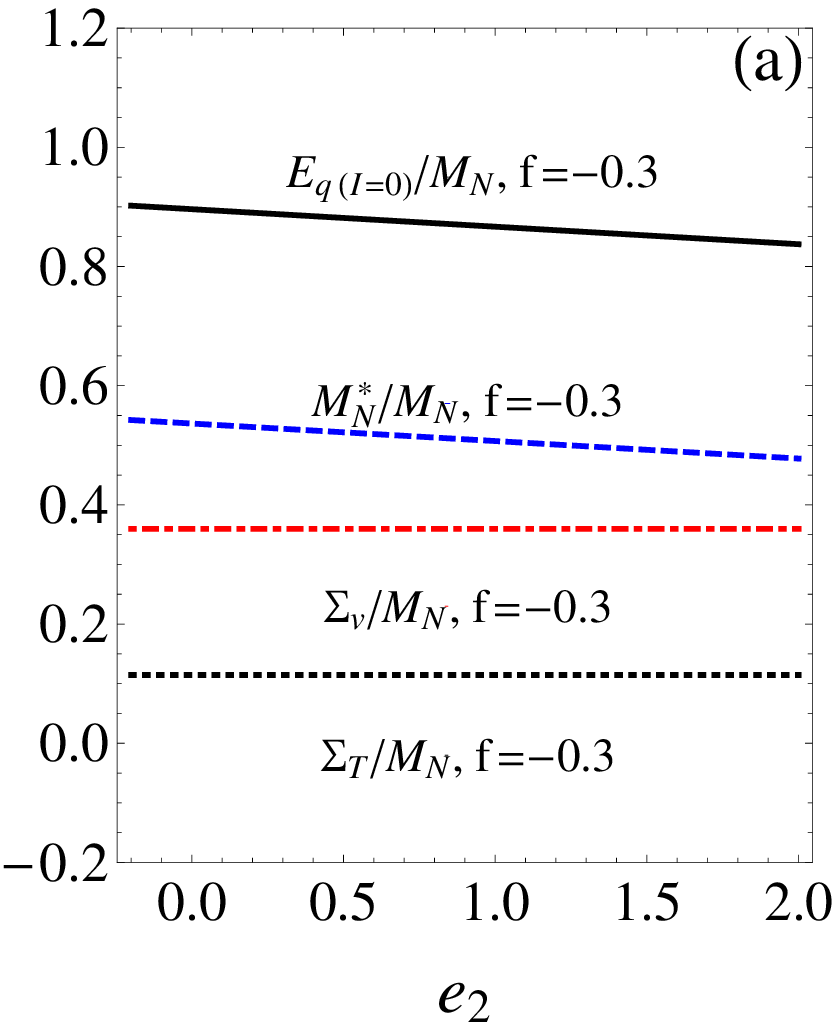}
\includegraphics[width=5cm]{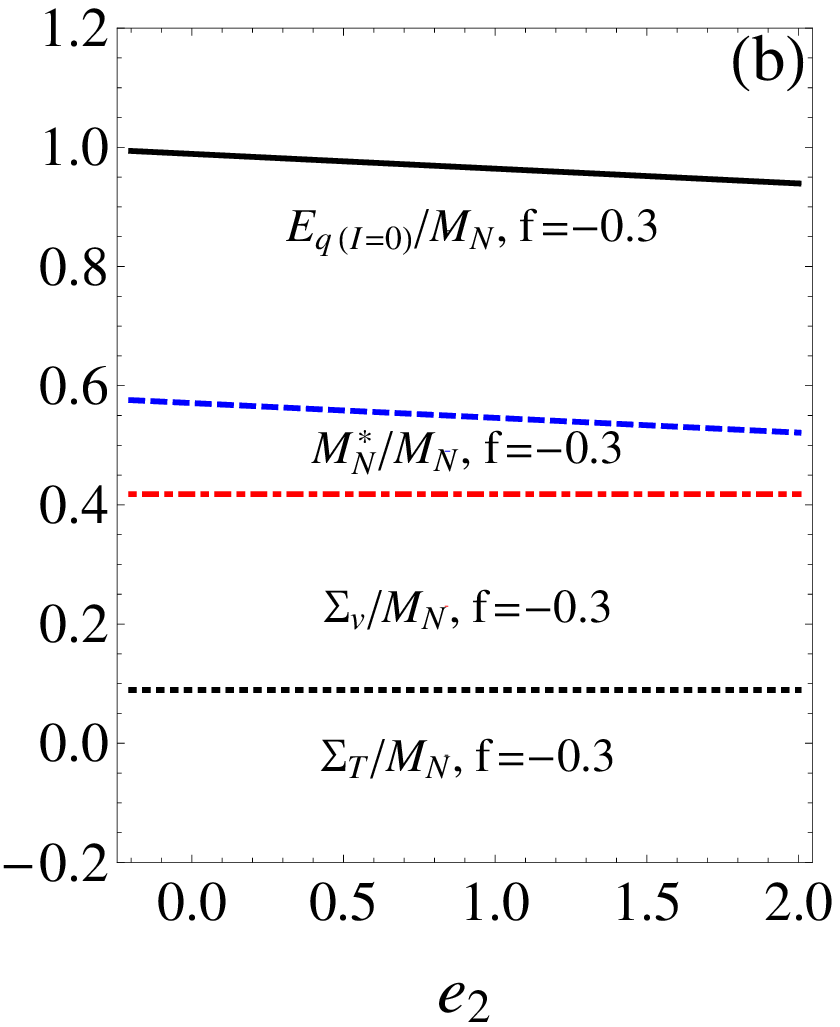}
\caption{(Color Online) The ratios between quasi nucleon self-energies and the vacuum mass as a function of $e_2$. 
(a) shows the results without 
and (b) with $\alpha_s$ corrections.} 
\label{fig:nmassfe}
\end{figure}

Next, we study the sum rule for the scalar ($M^*$),  vector self-energy ($\Sigma_V$), and 
the change in the pole mass [$E_q(I=0)$] for
symmetric nuclear matter, which can be obtained from combining the three invariant 
functions appearing in Eq.(\ref{eq:corr}) as discussed in \cite{Kie}. As can be seen in Fig. \ref{fig:nmassfe} (a), 
the  ratio $E_{q(I=0)}/M_N$ varies from 0.90 to 0.84 and
$M^{*}_N/M_N$  from 0.55 to 0.48, respectively,  as one changes
$e_2$ from $-0.20$ to $2.00$. On the other hand, when $\alpha_s$
corrections and dimension-eight condensates are included, the
corresponding values of $E_{q(I=0)}/M_N$ and $M^{*}_N/M_N$ are, 
respectively, modified to vary from 
1.00 to 0.93 and from 0.58 to 0.52 for the same
range of $e_2$. 
If $e_2$ can be restricted to the narrow range of 
Eq.(\ref{eq:new.estimate.e2}), this means that the dependence of both 
$E_{q(I=0)}/M_N$ and $M^{*}_N/M_N$ within this range clearly becomes very 
small. 
Moreover, previous sum rule calculations 
based on the larger value of $e_2$ given in 
Eq.(\ref{eq:DD2.200}) \cite{Furnstahl:1992pi}, 
have presumably overestimated the absolute changes of the above 
ratios in nuclear matter. 
\begin{figure}
\includegraphics[width=5cm]{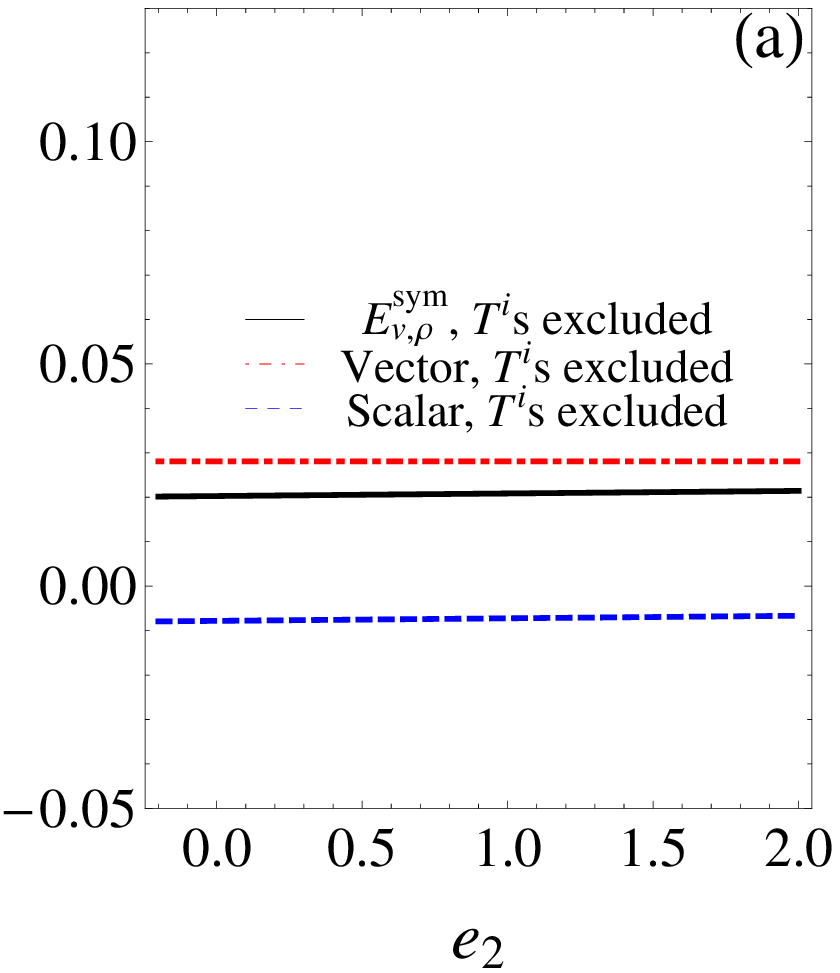}
\includegraphics[width=5cm]{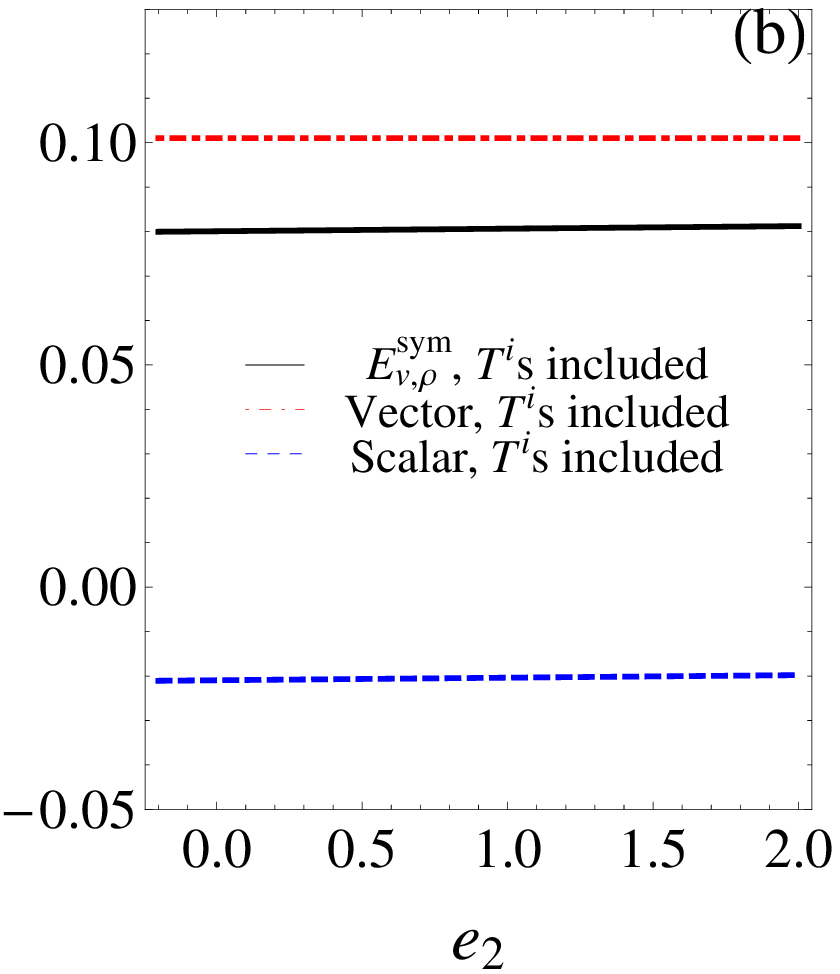}
\caption{(Color Online) Nuclear symmetry energy as a function of $e_2$ without (a) and with (b) twist-4 spin-2 four-quark matrix elements. 
The vertical axis is shown in units of GeV. 
For the detailed definition of the vector and scalar contributions, we refer the reader to \cite{Kie}.}
\label{fig:syme}
\end{figure}

Finally, let us check the modification of the symmetry energy as we change $e_2$ as above. 
In Fig. \ref{fig:syme}, we show the symmetry energy obtained with the full sum rule including $\alpha_s$ correction 
and dimension-eight condensate without (a) and with (b) the four-quark twist-4 spin-2 contributions. 
As can be seen in the figure, 
the curves are essentially constant for both cases and hence show no dependence on $e_2$. 
This simply shows that, 
in contrast to the nucleonic parameters shown in Fig. \ref{fig:nmassfe}, 
the contribution of $e_2$ to the symmetry energy is negligible.

\section{\label{Summary} Summary and Conclusion}
We have in this work studied the matrix element $\mathcal{S} \mathcal{T} \langle N| \overline{q} D_{\mu} D_{\nu} q |N \rangle$ and
have for the first time provided an estimate for its value that is based on experimental constraints.
To do this, we partially had to rely on a number of assumptions on the relative contributions of $u$ and $d$ quarks and their respective antiparticles
to $e_2$ and, furthermore, on the behavior of $e(x)$ at small and large $x$. For setting up these assumptions we followed the guidance
of the bag model \cite{Jaffe}, the chiral quark soliton model \cite{Ohnishi} and the spectator model \cite{Jakob}, which have been used to compute the
contributions of the different quark flavors to $e(x)$.

The final result, given in Eq.(\ref{eq:new.estimate.e2}), contains both the uncertainties of the experimental results of \cite{Courtoy} and
the systematic errors due to the assumptions derived from the different quark models.
We should stress here that, even though uncertainties of Eq.(\ref{eq:new.estimate.e2}) are large, the smallness of $e_2$ compared
to the earlier estimates of Eqs.(\ref{eq:old.est.1}) and (\ref{eq:DD2.200}) appears to be robust and does not depend on
the details of our employed assumptions. It hence seems to be essentially impossible to reconcile the experimental data of \cite{Courtoy} with
the old estimates of $e_2$, which, therefore, should be discarded in future studies.

To study the consequences of our findings, we have furthermore investigated the contribution of
$\mathcal{S} \mathcal{T} \langle N| \overline{q} D_{\mu} D_{\nu} q |N \rangle$ to the OPE relevant for deep inelastic scattering
and for the correlator of the vector meson current, that couples to the $\rho$, $\omega$ and $\phi$ meson states.
Our calculations show that, with the new estimate of Eq.(\ref{eq:new.estimate.e2}), $e_2$ is so small that its contribution to the
OPE for all the above-mentioned cases turns out to be negligible. Let us here especially mention the OPE corresponding to the $\phi$ meson
channel, for which the relevant operator is $m_s \mathcal{S} \mathcal{T} \langle N| \overline{q} D_{\mu} D_{\nu} q |N \rangle$.
With the old value of Eq.(\ref{eq:DD2.200}), this would have become the dominant twist-4 term, with seizable consequences for the behavior
of the $\phi$ meson in nuclear matter, but our result shows that this is not the case. 
We furthermore studied the contribution of the present 
operator to the nucleon scalar and vector self-energy and the nuclear symmetry energy in nuclear matter as obtained from the nucleon sum rule. 
From this, we found that the uncertainties of the self-energies due to $e_2$ are much reduced with our new estimate. 
For the nuclear symmetry energy, the effect of $e_2$ turned out to be minimal. 

Our new estimate of $e_2$ will also have consequences for the OPE of other channels not studied in this work, which will be
relevant for investigations of the behavior of the respective hadrons in nuclear matter. 
One such case could, for instance, be the study of the $D$ or $B$ meson spectrum at finite density \cite{Hilger,Zschocke}. 
As $e_2$, however, turns out to be small, it can generally be
expected to have only a marginal effect, but this needs to be confirmed in a separate calculation for each channel of interest.
The present authors have already started an investigation of the hyperon symmetry energy at finite density and found that it could 
indeed be non-negligible for this case \cite{Kie2}. 

\section*{Acknowledgments}
At an early stage of this work, P.G. 
was supported by the RIKEN Foreign Postdoctoral 
Researcher Program, the RIKEN iTHES Project. 
This work was supported 
by the Korean Research Foundation under Grants No. KRF-2011-0020333 and KRF-2011-0030621.

\end{document}